\documentclass[seceq,preprint]{ptptex}

\usepackage{graphicx}
\usepackage{url}

\notypesetlogo                       

\title{
Observables for identity-based tachyon vacuum solutions
}

\author{
Isao \textsc{Kishimoto}$^{1}$,
Toru \textsc{Masuda}$^{2}$
and Tomohiko
\textsc{Takahashi}$^{2}$
}

\inst{
$^1$Faculty of Education, Niigata University,
Niigata 950-2181, Japan\\
$^2$Department of Physics, Nara Women's University,
Nara 630-8506, Japan
}

\abst{
We consider a modified $KBc$ algebra in bosonic open string field theory
expanded around identity-based scalar solutions. By use of the algebra,
classical solutions on the background are constructed and observables
for them, including energy densities and gauge invariant overlaps, are
calculable. These results are applied to evaluate observables
analytically for both of the identity-based trivial pure gauge solution
and the identity-based tachyon vacuum solution.
}

\begin{document}
\maketitle
%

\section{Introduction
\label{sec:Introduction}
}

An analytic tachyon vacuum solution was constructed on the basis of the
identity string field, the BRST current, and the ghost field in bosonic
cubic open string field
theory.\cite{Takahashi:2002ez,Kishimoto:2002xi,Igarashi:2005wh}  
The identity string field is a fundamental object in the open string
field theory\cite{Witten:1985cc} and indeed it is a building block of
the $KBc$ algebra\cite{Okawa:2006vm} by which wedge-based
solutions\cite{Schnabl:2005gv} can be easily reconstructed. Then, the
identity-based solutions were found by some left-right splitting
algebra\cite{Takahashi:2002ez}, which is similar to the $KBc$ algebra in
a sense, and a certain type of the identity-based solutions can be
regarded as the tachyon vacuum solution. This is supported by evidence
from study of the theory expanded around the solution: vanishing
cohomology,\cite{Kishimoto:2002xi,Igarashi:2005wh,Inatomi:2011xr} 
no open string excitations,\cite{Takahashi:2003xe} and the existence of
the perturbative vacuum
solution.\cite{Kishimoto:2009nd,Kishimoto:2011zza} Hence, it appears
highly probable that observables for the identity-based tachyon vacuum
solution agree with those expected for the tachyon vacuum, although, due
to characteristic subtleties of the identity string field, it has been
difficult to perform direct evaluation of the observables.

Recently, significant progress has been made in the investigation of
identity-based marginal solutions. We have obtained a gauge equivalence
relation including the identity-based marginal solutions and some kind
of wedge-based tachyon vacuum solutions and, using this relation, we can
directly evaluate observables for the identify-based
solutions.\cite{Kishimoto:2013sra,Kishimoto:2014qza} The key ingredient
is a combined technique for the identity-based solutions and the $KBc$
algebra\cite{Inatomi:2012nv, Inatomi:2012nd} and it has potentiality for
investigating string field theory. In fact, it has been applied to
construct a new solution,\cite{Maccaferri:2014cpa} which has the same
algebraic structure as a wedge-based marginal
solution\cite{Kiermaier:2010cf} and is gauge equivalent to the
identity-based marginal solution. 

The main purpose of this paper is, based on these developments,
to confirm directly that the identity-based scalar 
solution provides the correct observables as expected.

The identity-based scalar solution is given by\cite{Takahashi:2002ez}
\begin{eqnarray}
 \Psi_0&=& Q_L(e^h-1)I-C_L((\partial h)^2 e^h)I,
\label{sol}
\end{eqnarray}
where $Q_L(f)$ and $C_L(f)$ are integrations of the BRST current
$j_{\rm B}(z)$ and the ghost $c(z)$,
which are multiplied by a function $f(z)$ along a
half unit circle. We find that the equation of motion holds for the
function $h(z)$ such that $h(-1/z)=h(z)$ and $h(\pm i)=0$. Moreover, the
reality condition of (\ref{sol}) imposes the function $h(z)$ to satisfy
$\left(h(z)\right)^*= h(1/z^*)$.

Expanding the string field $\Psi$ around the solution as
$\Psi=\Psi_0+\Phi$, we obtain an action for fluctuation:
\begin{eqnarray}
 S[\Psi;\,Q_{\rm B}]&=& S[\Psi_0;\,Q_{\rm B}]+ S[\Phi;\,Q'],
\end{eqnarray}
where we denote the action as
$S[\Psi;Q]=-\int \left(\frac{1}{2}
\Psi*Q\Psi+\frac{1}{3}\Psi*\Psi*\Psi\right)$
and the kinetic operator $Q'$ is given by
\begin{eqnarray}
 Q'&=& Q(e^h)-C((\partial h)^2 e^h).
\label{Qprime}
\end{eqnarray}
The operators $Q(f)$ and $C(f)$ are defined as integrations along a
whole unit circle.

We have a degree of freedom to choose a function $h(z)$ in the classical
solution and it can be changed by gauge transformations. Since
the function continuously connects to zero,
most of the solutions are regarded as a trivial pure gauge
solution. However, nontrivial solutions are generated at the boundary of
some function spaces. In the well studied case, the function includes
one parameter $a \geq -1/2$:
\begin{eqnarray}
 h_a(z)&=&\log\left(1+\frac{a}{2}(z+z^{-1})^2\right).
\label{haz}
\end{eqnarray}
It is known that the solution for $a>-1/2$ is a trivial pure gauge, but
it becomes a nontrivial solution for $a=-1/2$, 
for the reason mentioned above.

The transition from a trivial pure gauge to the tachyon vacuum
solution has been observed in various aspects
of the identity-based solution. In consequence, it is known that
zeros of $e^{h(z)}$ move on the $z$ plane with the deformation of
$h(z)$ and then the transition occurs when the zeros reach the
unit circle $|z|=1$. For example, $e^{h(z)}$ for (\ref{haz}) is
rewritten as
\begin{eqnarray}
 e^{h_a(z)}=\frac{1}{(1-Z(a))^2}\{z^2+Z(a)\}\{z^{-2}+Z(a)\}
~~\left(
Z(a)=\frac{1+a-\sqrt{1+2a}}{a}\right),
\label{haz2}
\end{eqnarray}
and it has zeros at $\pm \sqrt{-Z(a)}$ and
$\pm 1/\sqrt{-Z(a)}$.
When the parameter $a$ approaches $-1/2$ from
positive infinity,
$Z(a)$ runs from $1$ to $-1$ and then it takes the value $-1$ for
$a=-1/2$. As a result, we find that 
the zeros are on the unit circle only if $a=-1/2$ and then
the solution becomes the tachyon vacuum solution.\cite{Takahashi:2003xe}
For other functions, we find that the same transition occurs
if the zeros move to the unit circle.\cite{Igarashi:2005wh}

We now briefly outline our strategy.
First, we find the $KBc$ algebra in the shifted theory with the
action $S[\Phi;\,Q']$, which we call the $K'Bc$ algebra.
By means of the $K'Bc$ algebra, it is straightforward to construct
classical solutions in the shifted theory and it is possible to calculate
observables for these solutions.
Here, the shifted theory includes one
parameter $a$, as the above example, through $Q'$,  and so the
classical solutions depend on the parameter. This is similar to the case
of the analysis for the identity-based marginal
solutions,\cite{Kishimoto:2013sra,Kishimoto:2014qza} in which the
shifted theory and the
solution include parameters related to marginal deformations.
Therefore, according to the marginal case, we represent
the identity-based solution as a gauge equivalence
relation involving the identity-based and wedge-based solutions.
Finally, by use of this expression, we evaluate observables directly for
the identity-based solution.

Later we will see that there is a difference between the $K'Bc$ algebra
around the identity-based trivial solution 
and that around the identity-based nontrivial solution.
If $\Psi_0$ is a trivial pure gauge solution, the $K'Bc$
algebra can be transformed to the original $KBc$ algebra.
With the help of the transformation, we can calculate  
observables for classical solutions in the shifted theory.
Here, the existence of such a transformation depends crucially on the
positions of zeros of the function $e^{h(z)}$.
The zeros on the unit circle become obstacles to construction of the
transformation and therefore it is impossible to transform
from the $K'Bc$ algebra to the $KBc$ one in the case that $\Psi_0$ is the
tachyon vacuum solution. However, this implies
that, on the identity-based tachyon vacuum, $K'$, $B$, and $c$
have a different algebraic structure from the original $KBc$ algebra.
We will find that, on the identity-based tachyon vacuum, the operators
$K'$ and $c$ commute with each other and then all the solutions made of
$K'$, $B$ and 
$c$ can be written as modified BRST exact states. Accordingly,
observables for them are calculable even if $\Psi_0$ is the tachyon
vacuum solution.

This paper is organized as follows: First, we will consider
classical solutions in the theory expanded around the
identity-based solution in Sect.~\ref{sec:2}.
We construct the $K'Bc$ algebra with respect to $Q'$ and, by using the
$K'Bc$ algebra, we will find
classical solutions on the identity-based vacuum.
To calculate observables for the
classical solutions, we will construct a similarity transformation from
the operator $(K_1')_L$ to $(K_1)_L$. 
We will find that a conformal transformation is a significant part of
the similarity transformation and so we will illustrate
it by an example for $h_a(z)$ in (\ref{haz}).
Then, we will calculate observables for the classical solutions 
around $\Psi_0$.
In Sect.~\ref{sec:3},
based on the results in the previous section, we analytically evaluate
observables for the identity-based solutions. In Sect.~\ref{sec:4}, 
we will give concluding remarks.
In Appendix~\ref{sec:appendixA}, we provide a detailed proof of the
properties of a differential equation that plays an important role
in the calculation of observables.\\

\noindent{\it Note added:}
When we had a discussion with N.~Ishibashi during the conference
SFT2014 at SISSA, Trieste, it was found that we had reached the same
conclusion for the gauge invariant observables for the identity-based
tachyon vacuum solution.\footnote{
Both Ishibashi's and our results were presented independently at the
conference.
The presentation files of these talks by N.~Ishibashi and one of the
authors (T.T.) are available on the conference website:
\url{http://www.sissa.it/tpp/activity/conferences/SFT2014/}.
}
 The main difference is that
he argued in detail for a regularization method to evaluate the
observables\cite{Ishibashi}
but we evaluated the observables for the identity-based trivial solution
in addition to the tachyon vacuum case.

After almost completing the manuscript, we found a paper by
S.~Zeze,\cite{Zeze:2014qha} which treats similar
problems with different methods.

\section{Classical solutions around the identity-based solution
\label{sec:2}
} 

\subsection{Modified $KBc$ algebra
\label{sec:2-1}}

We can construct a modified $KBc$ algebra associated with the deformed
BRST operator (\ref{Qprime}):
\begin{eqnarray}
&&
 K'=Q'B,\ \ Q'K'=0,\ \ Q'c=cK'c,
\label{K'Bc}\\
&&
 B^2=0,\ \ c^2=0,\ \ Bc+cB=1,
\label{K'Bc2}
\end{eqnarray}
where $B$ and $c$ are the same string fields in the conventional $KBc$
algebra,\cite{Okawa:2006vm} and $K'$ is given by\footnote{
We use the following convention:
\begin{align}
&B=\frac{\pi}{2}(B_1)_LI,~~~c=\frac{1}{\pi}c(1)I,~~~
(B_1)_L=\int_{C_{\rm left}}\frac{dz}{2\pi i}(1+z^2)b(z),
\end{align}
where $I$ is the identity string field and
the integration path $C_{\rm left}$ on the $z$-plane is a half unit circle:
$|z|=1,~~{\rm Re}\,z\ge 0$. 
}
\begin{eqnarray}
&&
 K'= \frac{\pi}{2}(K_1')_L I,\ \ \ (K_1')_L=
\{Q',\,(B_1)_L\}.
\end{eqnarray}
The operator $(K_1')_L$ is explicitly calculated as
\begin{eqnarray}
 (K_1')_L
=\int_{C_{\rm left}} \frac{dz}{2\pi i}
(1+z^2)\left\{
e^{h(z)}T(z)+(\partial h)e^{h(z)}j_{\rm gh}(z)+
\left(\frac{3}{2}\partial^2 h+\frac{1}{2}(\partial h)^2\right)
e^{h(z)}\right\},~~
\label{K1primeL}
\end{eqnarray}
where $j_{\rm gh}(z)$
is the ghost number current and $T(z)$ is the total energy momentum
tensor.\footnote{It can be calculated
by using the relations in Ref.~\citen{Inatomi:2011xr}:
\begin{eqnarray}
 \{Q(f),\,b(z)\}=\frac{3}{2}\partial^2 f(z)+\partial f(z) j_{\rm
  gh}(z)
+f(z)T(z),\ \ \ 
\{C(f),\,b(z)\}= f(z).\nonumber
\end{eqnarray}
}
We can easily find that if $h(z)$ becomes identically zero, 
the operator $(K'_1)_L$ is equal to the conventional $(K_1)_L$ in the
$KBc$ algebra. 

For the general function $h(z)$,
$K'$, $B$, $c$, and $Q'$ have the same
algebraic structure as that of the $KBc$ algebra.
However, if we choose a special function, the algebra is
more simplified.
To see this let us consider the relation $Q'c=cK'c$ in
(\ref{K'Bc}). This relation is derived from the following equations:
\begin{eqnarray}
 \{Q(e^h),\,c(z)\}&=& e^{h(z)}c\partial c(z),\label{Qehc}\\
 \left[ K_1',\,c(z) \right]&=& -(\partial (1+z^2))e^{h(z)}c(z)+
(1+z^2)e^{h(z)}\partial c(z),
\label{K1c}
\end{eqnarray}
where $K_1'$ is the operator defined by the replacement of the
integration path in (\ref{K1primeL}) with a unit circle:
$K_1'=\{Q',B_1\}=\{Q',b_1+b_{-1}\}$. 
As mentioned 
in Sect.~\ref{sec:Introduction}, 
the function $e^{h(z)}$ has zeros on the unit circle
in the case that the solution becomes the tachyon vacuum solution.
Indeed, for (\ref{haz}), $e^{h_a(z)}$ has zeros at $z=\pm 1$ only
in the case $a=-1/2$ and then,
from (\ref{Qehc}), $Q(e^{h_a})$ and $c(1)$ anticommute with
each other for $a=-1/2$.\footnote{
We note that
$e^{-i\sigma}c(e^{i\sigma})I=-e^{-i(\pi-\sigma)}c(e^{i(\pi-\sigma)})I$
and therefore $c(1)I=c(-1)I$.
}
Similarly, $K_1'$ and $c(1)$
commute for $a=-1/2$ from (\ref{K1c}).
Consequently, we find a simplified algebra only in the case that
$e^{h(z)}$ has zeros at $z=\pm 1$, namely in the theory around the
tachyon vacuum solution:
 \begin{eqnarray}
&&
 K'=Q'B,\ \ Q'K'=0,\ \ Q'c=0,
\label{sK'Bc}\\
&&
 B^2=0,\ \ c^2=0,\ \ Bc+cB=1.
 \label{sK'Bc2}
\end{eqnarray}

Actually, there are other possibilities\cite{Kishimoto:2002xi} where the
function $e^{h(z)}$
for the tachyon vacuum solution has zeros on the unit circle 
but not at $z=\pm 1$.
We will discuss these cases at the end of the section.

\subsection{Classical solutions
\label{sec:2-2}
}

The equation of motion in the theory around the solution \eqref{sol}
is given by
\begin{eqnarray}
 Q'\Phi+\Phi^2=0,
\label{eom}
\end{eqnarray}
where $Q'$ is the modified BRST operator (\ref{Qprime}).
We can find various classical solutions in the shifted background by
substituting $K'$ for $K$ in the solutions given by the $KBc$ algebra in
the original theory. In the conventional theory with $Q_{\rm B}$, a
classical solution using the $KBc$ algebra is written as
\begin{eqnarray}
 \Psi_0(K,B,c)
&=& \sum_{ij}{\cal A}_i(K)\,c\,{\cal B}_j(K)+\sum_{ijk}
{\cal C}_i(K)\,c\,{\cal D}_j(K)\,c\,
{\cal E}_k(K)B,
\label{originalsol}
\end{eqnarray}
which is general configuration with ghost number one in terms of the
$KBc$ algebra. Here, ${\cal A}_i(K),~{\cal B}_i(K),~{\cal C}_i(K),~{\cal
D}_i(K)$, and ${\cal E}_i(K)$ are appropriate functions of the string
field $K$. Once a particular solution (\ref{originalsol}) is given, a
classical solution for (\ref{eom}) is constructed as 
\begin{eqnarray}
 \Phi_0(K',B,c)
&=& \sum_{ij}{\cal A}_i(K')\,c\,{\cal B}_j(K')
+\sum_{ijk}{\cal C}_i(K')\,c\,{\cal D}_j(K')\,c\,
{\cal E}_k(K')B.
\label{solKprime}
\end{eqnarray}

If $\Psi_0(K,B,c)$ is a solution in the conventional theory,
$\Phi_0(K',B,c)$ is a solution in the shifted background, regardless of
whether or not the $K'Bc$ algebra is simplified as (\ref{sK'Bc}).
However, we emphasize that in the case that the algebra is simplified,
the solution has a simpler expression:
\begin{eqnarray}
 \Phi_0(K',c)={\cal F}(K')c,
\label{solKprime2}
\end{eqnarray}
where ${\cal F}(K')=\sum_{ij}{\cal A}_i(K'){\cal B}_j(K')$ and the
second term in (\ref{solKprime}) vanishes due to $K'c=cK'$ and $c^2=0$.

\subsection{Transformations from $(K_1')_L$ to $(K_1)_L$
\label{sec:2-3}
}
In this subsection, we will consider a similarity transformation from
$(K'_1)_L$ to the conventional $(K_1)_L$. 

First, we introduce the
operator\cite{Takahashi:2002ez,Takahashi:2003xe}\footnote{
This operator was written as $K(h)$ in
Ref.~\citen{Takahashi:2003xe}. 
The ghost number current $j_{\rm gh}(z)$ is defined by using
$SL(2,{\mathbb R})$ normal ordering.
If $h(z)$ satisfies $h(-1/z)=h(z)$,
owing to the second term ($-3/2 z^{-1}$), the operator
is transformed as $\tilde{q}(h)\rightarrow -\tilde{q}(h)$ under the
BPZ conjugation. Moreover, $\tilde{q}(h)$ is a derivation 
with respect to the star product among string fields.
}
\begin{eqnarray}
 \tilde{q}(h)&=&\oint \frac{dz}{2\pi i} h(z)
\left(j_{\rm gh}(z)-\frac{3}{2}z^{-1}\right),
\end{eqnarray}
where $j_{\rm gh}(z)$ is the ghost number current, $j_{\rm gh}=cb$.
Using this operator, the modified BRST operator (\ref{Qprime}) is
transformed to the original BRST
operator:\cite{Takahashi:2002ez}\footnote{
The operator $e^{\pm \tilde{q}(h)}$ becomes singular for the tachyon
vacuum solution.\cite{Takahashi:2002ez}
}
\begin{eqnarray}
 e^{-\tilde{q}(h)}Q'e^{\tilde{q}(h)}=Q_{\rm B}.
\label{Qtrans}
\end{eqnarray}
Accordingly, we can remove the ghost number current from $(K_1')_L$ in
(\ref{K1primeL}) by a similarity transformation:
\begin{eqnarray}
 e^{-\tilde{q}(h)}(K_1')_L\,e^{\tilde{q}(h)}&=&
 e^{-\tilde{q}(h)}\left\{Q',\,(B_1)_L\right\}\,e^{\tilde{q}(h)}
= 
\left\{Q_{\rm B},\,e^{-\tilde{q}(h)}(B_1)_L\,e^{\tilde{q}(h)}
\right\}\nonumber\\
&=&
\int_{C_{\rm left}}\frac{dz}{2\pi i}(1+z^2)e^{h(z)}T(z),
\label{ehT}
\end{eqnarray}
where we have used
$e^{-\tilde{q}(h)}b(z)e^{\tilde{q}(h)}
=e^{h(z)}b(z)$.\cite{Takahashi:2003xe}.

Next, we look for a conformal transformation $z'=f(z)$ that maps
 (\ref{ehT}) to $(K_1)_L$. Since $T(z)$ is a primary field
with the dimension 2, 
the operator (\ref{ehT}) is transformed as
\begin{eqnarray}
f\left[\int_{C_{\rm left}}\frac{dz}{2\pi i}(1+z^2)e^{h(z)}T(z)
\right]&=&
 \int_{C'_{\rm left}}\frac{df}{2\pi i}(1+z^2)e^{h(z)}\frac{df}{dz}
T(f(z)),
\label{f(ehT)}
\end{eqnarray}
where $C'_{\rm left}$ is an integration path in the mapped plane
such as $f:C_{\rm left}\to C'_{\rm left}$.
In order that (\ref{f(ehT)}) coincides with $(K_1)_L$,
the function $f(z)$ must satisfy a differential equation:
\begin{eqnarray}
 (1+z^2)e^{h{(z)}}\,\frac{df}{dz}&=& 1+f^2,
\label{deqf}
\end{eqnarray}
and $C'_{\rm left}$ must remain the same path along the left
half of a string. 

To find the conformal map, it is necessary to solve the differential
equation (\ref{deqf}) in an annulus including the unit circle
$|z|=1$. The important point is that we can solve it if
$e^{h(z)}$ has no zeros on the unit circle, as seen in
Appendix~\ref{sec:appendixA}. Moreover,
we can prove that under the initial condition $f(1)=1$,
the solution $f(z)$ has the following properties:
\begin{eqnarray}
1.~~~~ &&|z|=1\ \Rightarrow\ |f(z)|=1,
\label{prop1}\\
2.~~~~ &&f:\ C_{\rm left}\ \twoheadrightarrow\ C_{\rm left},
\label{prop2}\\
3.~~~~ &&f\left(-\frac{1}{z}\right)=-\frac{1}{f(z)}.
\label{prop3}
\end{eqnarray}
We illustrate these by the solution given for
$h_a(z)$ (\ref{haz}) in the next subsection and
we give a detailed proof in Appendix~\ref{sec:appendixA}.

From (\ref{prop1}) and (\ref{prop2}), 
we find that the conformal map by the solution $f$ 
leaves the integration path in (\ref{f(ehT)}) unchanged, namely
$C'_{\rm left}=C_{\rm left}$. 
Therefore we can transform the operator (\ref{ehT}) to $(K_1)_L$ by the
conformal transformation $f$.
Moreover, (\ref{prop3})
indicates that the conformal map $f(z)$ is generated by
the operators $K_n=L_n-(-1)^n L_{-n}$.\footnote{$K_n$ generates a
transformation 
$f(\sigma)$ such that $f(\pi-\sigma)=\pi-f(\sigma)$.\cite{Witten:1985cc}
By setting 
$z=e^{i\sigma}$, this corresponds to (\ref{prop3}).} 

Consequently,
around the trivial pure gauge solution, we can construct 
the similarity transformation
\begin{eqnarray}
U_f e^{-\tilde{q}(h)}(K_1')_L\,e^{\tilde{q}(h)}U_f^{-1}&=&
(K_1)_L,
\label{UfEqK}
\end{eqnarray}
where $U_f$ is the operator for the conformal transformation $f$
and it is given in the form
\begin{eqnarray}
 U_f=\exp\left(\sum_n v_n K_n\right),
\label{Uf}
\end{eqnarray}
with certain parameters $v_n$.

For the identity-based tachyon vacuum solution,
a solution to the differential equation (\ref{deqf}) for $f(z)$
has singularity due to zeros of $e^{h(z)}$ on the unit circle 
(Appendix \ref{sec:appendixA}). In this
sense, we emphasize that a regular operator $U_f$ 
does not exist for the tachyon vacuum.

\subsection{An example for the transformation
\label{secexample}}

We illustrate the existence of the transformation $U_f$
by solving (\ref{deqf}) for (\ref{haz}).
For (\ref{haz}), $e^{h(z)}$ is written as (\ref{haz2}),
and, under the initial condition
$f(1)=1$, setting $z=e^{i\sigma}$, we can solve the
differential equation (\ref{deqf}) as follows: 
\begin{eqnarray}
 f(e^{i\sigma})&=& e^{i\phi(\sigma)},\ \ 
\phi(\sigma)=\sigma+2\arctan \frac{g(\sigma) \cos\sigma}{
1+g(\sigma)\sin\sigma},
\label{fsigma}
\end{eqnarray}
where, for $-1/2<a\leq 0\ (-1<Z(a)\leq 0)$, $g(\sigma)$ is given as
\begin{eqnarray}
 g(\sigma)&=& \tanh\left\{
\frac{\sqrt{-Z(a)}}{1+Z(a)}\arctan\left(\frac{2\sqrt{-Z(a)}}{1+Z(a)}
\sin\sigma\right)\right\},
\end{eqnarray}
and, for $0<a\ (0<Z(a)<1)$, it is
\begin{eqnarray}
 g(\sigma)&=& -\tanh\left\{
\frac{\sqrt{Z(a)}}{1+Z(a)}{\rm arctanh} \left(\frac{2\sqrt{Z(a)}}{1+Z(a)}
\sin\sigma\right)\right\}.
\end{eqnarray}

Since $\phi(\sigma)$ is a real-valued function for $\sigma\in {\mathbb R}$, 
the solution (\ref{fsigma}) satisfies (\ref{prop1}). 
By differentiating $\phi(\sigma)$, 
it can be seen that
$\phi(\sigma)$ is a monotonically
increasing function for $-\pi/2<\sigma<\pi/2$.
We also see that $\phi(\pm \pi/2)=\pm \pi/2$.
Therefore, $C'_{\rm left}$ is the same as the
left half of a string, then (\ref{prop2}) is satisfied. 
Moreover, since $g(\pi-\sigma)=g(\sigma)$, we find that
$\phi(\pi-\sigma)=\pi-\phi(\sigma)$ and then
the function (\ref{fsigma}) satisfies (\ref{prop3}).
Thus, the solution (\ref{fsigma})
satisfies (\ref{prop1}), (\ref{prop2}) and (\ref{prop3})
in the case of $a>-1/2$, and then the transformation $U_f$ exists.

Here, we should emphasize that
the transformation (\ref{Uf}) 
exists only in the case $a>-1/2$ and
it does not at $a=-1/2$,
because the circle-to-circle correspondence for the integration path
is broken down for $a=-1/2$.
In fact, taking the limit $a\rightarrow -1/2$,
the phase $\phi(\sigma)$ in (\ref{fsigma}) approaches a step function:
\begin{eqnarray}
 \lim_{a\rightarrow -1/2}\phi(\sigma)&=&
\left\{\begin{array}{ll}
 \displaystyle
\frac{\pi}{2}& (0<\sigma<\pi)\\
&\\
 \displaystyle
-\frac{\pi}{2}& (-\pi<\sigma<0)
       \end{array}\right..
\end{eqnarray}
Therefore, we cannot transform $(K'_1)_L$ to $(K_1)_L$ by a regular
conformal map at $a=-1/2$.

\subsection{Observables around the trivial pure gauge solution
\label{sec:2-5}
}

In this subsection, we will show that observables for the solution
(\ref{solKprime}) around the identity-based trivial pure gauge solution
are equivalent to those for the original solution (\ref{originalsol}).

First, we find that
$c(1)$ and $(B_1)_L$ are invariant under the similarity transformation
(\ref{UfEqK}),
namely, $U_f e^{-\tilde{q}(h)}c(1)e^{\tilde{q}(h)}U_f^{-1}=
c(1)$ and $U_f e^{-\tilde{q}(h)}(B_1)_L\,e^{\tilde{q}(h)}U_f^{-1}=
(B_1)_L$.
Using 
$e^{-\tilde{q}(h)}c(z)e^{\tilde{q}(h)}
=e^{-h(z)}c(z)$\cite{Takahashi:2003xe}, we have
\begin{eqnarray}
 U_f e^{-\tilde{q}(h)}c(z)e^{\tilde{q}(h)}U_f^{-1}
=e^{-h(z)}\left(\frac{df(z)}{dz}\right)^{-1}
c(f(z)).
\end{eqnarray}
From the differential equation (\ref{deqf}), it follows that
\begin{eqnarray}
 =\frac{1+z^2}{1+f(z)^2}\,c(f(z)),
\end{eqnarray}
and then $c(1)$ is invariant under the transformation because $f(1)=1$
is imposed as the initial condition. With regard to $(B_1)_L$, the
invariance can be easily seen by using
$e^{-\tilde{q}(h)}b(z)e^{\tilde{q}(h)}=e^{h(z)}b(z)$\cite{Takahashi:2003xe}
and the fact that $b(z)$ is a primary field with the dimension 2.

Now that the similarity transformation of $(K'_1)_L$, $(B_1)_L$,
and $c(1)$ is established, we can transform the solution
(\ref{solKprime}) to the original solution (\ref{originalsol}).
An important point is that the generators $\tilde{q}(h)$
and $K_n$ are derivations with respect to the star product
and in particular $\tilde{q}(h)I=0$
and $K_nI=0$. Then, we obtain the transformation from 
string fields $(K,B,c)$ to $(K',B,c)$:
\begin{eqnarray}
K'=e^{\tilde{q}(h)}U_f^{-1} K,~~
B=e^{\tilde{q}(h)}U_f^{-1} B,~~
c=e^{\tilde{q}(h)}U_f^{-1} c.
\label{KBctrans}
\end{eqnarray}
Noting that $U_f^{-1}$ and $e^{\tilde{q}(h)}$ are given as an
exponential of derivations, we find that the solution (\ref{solKprime})
is given as a transformation from (\ref{originalsol}):
\begin{eqnarray}
 \Phi_0(K',B,c)=e^{\tilde{q}(h)}\,U_f^{-1}\Psi_0(K,B,c).
\label{soltrans}
\end{eqnarray}

Let us consider the vacuum energy for $\Phi_0$ around the
trivial pure gauge solution. Using the transformation (\ref{soltrans}),
the action for $\Phi_0$ is given by
\begin{eqnarray}
 S[\Phi_0(K',B,c);\,Q']=S[\Psi_0(K,B,c);\,U_f Q_{\rm B} U_f^{-1}],
\end{eqnarray}
where we have used (\ref{Qtrans})
and the BPZ conjugation: $e^{\tilde{q}(h)}U_f^{-1} \to U_fe^{-\tilde{q}(h)}$. 
Since $U_f$ is generated by
$K_n$ and the operators
$Q_{\rm B}$ and $L_n$ commute with each other,
$U_f Q_{\rm B} U_f^{-1}$ is equal to $Q_{\rm B}$. As a result,
the vacuum energy for
$\Phi_0(K',B,c)$ is equivalent to that for $\Psi_0(K,B,c)$
in the conventional theory.\\

Next,
let us consider gauge invariant overlaps for $\Phi_0(K',B,c)$.
The gauge invariant overlap
for the open string field $\Psi$ 
is defined as\cite{Zwiebach:1992bw}
\begin{eqnarray}
 O_V(\Psi)=\langle I|V(i)|\Psi\rangle,
\end{eqnarray}
where $V(i)$ is a closed string vertex operator,
such as $c(i)c(-i)V_{\rm m}(i,-i)$, where $V_{\rm m}(z,\bar{z})$
is a matter primary with the conformal dimension $(1,1)$.
 Noting that $h(\pm i)=0$,
in spite of the closed string vertex on $I$, 
$\tilde{q}(h)$ satisfies
\begin{eqnarray}
 \langle I|V(i)\,\tilde{q}(h)=0.
\end{eqnarray}
In addition, the operators $K_n$ generate a global symmetry
of the open string field theory even if the gauge invariant overlaps are
introduced as sources.
In fact, since $V(i)$ has the dimension $0$ and
$f(\pm i)=\pm i$, we find that $U_f V(i)
U_f^{-1}=V(f(i))=V(i)$ and then
\begin{eqnarray}
 \langle I|V(i)\,U_f^{-1} =\langle I|V(i).
\end{eqnarray}
Consequently,
we can see that the gauge invariant overlaps for
the solution (\ref{solKprime}) are equivalent to that for the
conventional solution (\ref{originalsol}):\cite{Ellwood:2008jh,Kawano:2008ry}
\begin{eqnarray}
 O_V(\Phi_0(K',B,c))=O_V(\Psi_0(K,B,c)).
\end{eqnarray}

\subsection{Observables around the tachyon vacuum solution
\label{sec:ObsTV}}

Now let us consider observables for the classical solution (\ref{solKprime2})
around the identity-based tachyon vacuum solution.
In this vacuum, the modified $K'Bc$ algebra and the classical solution
are simplified as mentioned before,
and $Q'$ has vanishing cohomology
\cite{Kishimoto:2002xi,Inatomi:2011xr}.
From $Q'c = 0$ in
(\ref{sK'Bc}), $c$ turns out to be an exact state
with respect to the modified BRST operator $Q'$.
Since $Q'K'=0$,
the solution (\ref{solKprime2}) can be written as a
modified BRST exact state:
\begin{eqnarray}
 \Phi_0(K',c)&=& Q'\chi.
\label{PhiQchi}
\end{eqnarray}
Therefore, we conclude that both the vacuum energy and the gauge
invariant overlaps are zero for the classical solution
(\ref{solKprime2}).

Here, we should note that the derivation of (\ref{PhiQchi}) requires
careful consideration. In Ref.~\citen{Inatomi:2011xr}, the homotopy
operator is given for the BRST operator $Q'$ in the identity-based
tachyon vacuum. In the case of the tachyon vacuum solution using the
function (\ref{haz}) with $a=-1/2$, 
a corresponding homotopy operator is $\hat{A}=(b(1)+b(-1))/2$.
If it is used for the above exact form, such as
$\chi={\cal F}(K')\hat{A}c$, the divergence arises from a collision
between $b(1)$ and $c(1)$ and so we need regularization of
(\ref{solKprime2}). However, it should be noted 
that the homotopy operator $\hat{A}$, such as $\{\hat{A},Q'\}=1$, is not
unique because we can add to it a commutator $[Q', {\cal O}]$, where
${\cal O}$ is an arbitrary operator with ghost number $-2$. Then,
we have the possibility of providing a regularization procedure by adding
such terms to the homotopy operator. In Ref.~\citen{Ishibashi}, several
regularization methods are rigorously discussed.

In addition, we should notice that the divergence does not appear in
the procedure used to analyze the cohomology of $Q'$
in~Ref.~\citen{Kishimoto:2002xi}.
The string field $c$ can be expanded in the Fock space
for each $L_0$-level and the lowest-level state in $c$ is
$c_1\left|0\right>$: 
\begin{eqnarray}
 c=\frac{1}{2\pi}c_1\left|0\right>+\cdots.
\end{eqnarray}
For the identity-based tachyon vacuum using the function (\ref{haz})
with $a=-1/2$,  $Q'$ has an oscillator expression:
\begin{eqnarray}
 Q'=R_2+R_0+R_{-2},
\end{eqnarray}
where $R_n$ stands for terms with the mode number $n$ with respect to
$L_0$:\footnote{ 
Here, we have expanded the conventional primary BRST current $j_{\rm B}$
as $j_{\rm B}(z)=\sum_{n=-\infty}^{\infty}Q_n z^{-n-1}$
and therefore $Q_0=Q_{\rm B}$ in particular. The nilpotency of $Q'$
leads to the anticommutation relations, 
\begin{eqnarray*}
\{R_{\pm 2},\,R_{\pm 2}\}=0,\ \{R_{\pm 2},\,R_0\}=0,\ \ 
2\{R_2,\,R_{-2}\}+\{R_0,R_0\}=0.
\end{eqnarray*}
}
\begin{align}
&R_{\pm 2}=-\frac{1}{4}Q_{\pm 2}+c_{\pm 2},
&R_0=\frac{1}{2}Q_{\rm B}+2c_0.
\end{align}
Then,
$Q'c$ is written by the Fock space state starting from a lowest-level
state: 
\begin{eqnarray}
 Q'c=\frac{1}{2\pi}R_2c_1\left|0\right>+\cdots.
\end{eqnarray}
Therefore, we can solve the equation $Q'c=0$ level by level and then $c$
can be given by a modified BRST exact state with no divergence,
because, as in~Ref.~\citen{Kishimoto:2002xi}, the cohomology of $Q'$
is expressed by the well-defined Fock space expression. In particular, 
there is no cohomology within the ghost number one sector.
Thus, by solving the cohomology level by level, we can write $c$ as a
$Q'$ exact state with a well-defined Fock space expression. As a result,
the expression (\ref{PhiQchi}) can be well-defined with no divergence.

\subsection{Comments on a simplified algebra
\label{sec:2-7}
}

Here, we comment on the case that $e^{h(z)}$ for the 
identity-based solution has zeros on the unit circle 
but not at $z=\pm 1$.
In Ref.~\citen{Kishimoto:2002xi},  the identity-based solutions
(\ref{sol}) with the function $h^l_a(z)$ ($l=1,2,3,\cdots$; $a\ge
-1/2$): 
\begin{align}
h^l_a(z)&=\log\left(1-\frac{a}{2}(-1)^l
\left(z^l-(-z^{-1})^l\right)^2\right) 
\end{align}
were considered as a generalization of the function $h_a(z)$
(\ref{haz}), which is the case of $l=1$ in the above. The solution
corresponding to $h_a^l(z)$ is pure gauge for $a>-1/2$ and we can apply
the prescriptions in the previous subsections. In the case that
$a=-1/2$, the corresponding solution is believed to represent the
tachyon vacuum, where the BRST operator $Q'$ around the solution has no
cohomology,\cite{Kishimoto:2002xi, Inatomi:2011xr} and  we have
\begin{align}
e^{h_{-1/2}^l(z)}&=\frac{(-1)^l}{4}(z^l+(-z^{-1})^l)^2.
\label{ehl-1/2}
\end{align}
It has zeros at $z=\pm 1$ when  $l$ is a positive odd integer
and we can use simplified algebra (\ref{sK'Bc}) in the same way as the case of
the function (\ref{haz}).
 
In the case that $l=2m$ ($m=1,2,\cdots,$), i.e. a positive even
integer, the function (\ref{ehl-1/2})  
has zeros on the unit circle: $z_k=e^{i\theta_k}$, where
$\theta_k=\frac{2k-1}{4m}\pi$ 
($k=1,2,\cdots,4m$), and they are not $\pm 1$.
In this case, the simplified $K'Bc$ algebra (\ref{sK'Bc}) does not hold
because $e^{h_{-1/2}^{2m}(1)}\ne 0$. However, we can obtain a simplified
algebra by using 
\begin{align}
c'&=\frac{1}{\pi\cos\theta_1}e^{-i\theta_1}c(e^{i\theta_1})I
\label{c'def}
\end{align}
with $\theta_1=\frac{\pi}{4m}$ instead of $c=\frac{1}{\pi}c(1)I$
as follows. Firstly, we note that 
\begin{align}
&e^{\alpha K_1}c=\frac{2}{\pi}U_1^{\dagger}U_1\tilde c(\alpha)|0\rangle,
\end{align}
where $\tilde c(\tilde z)=\tan\circ\,c(\tilde z)$ and we have used the
notation in Ref.~\citen{Schnabl:2005gv}. 
Using a relation
\begin{align}
&(U_1^{\dagger}U_1)^{-1}c(e^{i\theta})(U_1^{\dagger}U_1)
=\left(\cos(it +\frac{\pi}{4})\right)^{-2}\tilde c(it),
&e^{i\theta}=\tan(it+\frac{\pi}{4}),
\end{align}
we have
\begin{align}
e^{itK_1}c=\frac{2}{\pi}\cos^2(it+\frac{\pi}{4})
c(e^{i\theta})I=\frac{1}{\pi\cos\theta}e^{-i\theta}c(e^{i\theta})I.
\end{align}
Therefore, with $t_1$ such as $e^{i\theta_1}=\tan(it_1+\frac{\pi}{4})$,
or $t_1={\rm arctanh}(\tan\frac{\theta_1}{2})$,
$c'$ defined in (\ref{c'def}) can be expressed as $c'=e^{it_1K_1}c$.
Because $K_1=L_1+L_{-1}$ is a derivation with respect to the star
product, and noting $[K_1,(B_1)_L]=0$ and $[K_1,Q_{\rm B}]=0$, we have
\begin{align}
&Q_{\rm B}c'=e^{it_1K_1}Q_{\rm B}c=c'Kc',
&&Bc'+c'B=e^{it_1K_1}(Bc+cB)=1,
&&(c')^2=e^{it_1K_1}c^2=0,
\end{align}
and they form a kind of $KBc'$ algebra.
Furthermore, noting (\ref{Qehc}) and (\ref{K1c}), we obtain a simplified
algebra 
\begin{align}
&K'=Q'B,~~~Q'K'=0,~~~Q'c'=0,\\
&B^2=0,~~~(c')^2=0,~~~Bc'+c'B=1,
\end{align}
for a modified BRST operator $Q'$ corresponding to the function
$h_{-1/2}^{2m}(z)$.  
Using the above, we can apply the prescription in Sect.~\ref{sec:ObsTV}
in a similar way. 

\section{Observables for identity-based solutions
\label{sec:3}
}

We consider direct calculation of observables for the identity-based
tachyon vacuum solutions, by use of the method for the identity-based
marginal solution in Ref.~\citen{Kishimoto:2013sra}.

We consider one parameter family of the identity-based
solution, $\Psi_0(a)$. The parameter $a$ deforms the
function $h(z)$ in 
the solution, and as the simplest case (\ref{haz}) it takes values
$a\geq -1/2$, the solution becomes the tachyon vacuum at
$a=-1/2$,\footnote{Namely, $e^{h(z)}$ has zeros on the unit circle at
$a=-1/2$. } otherwise it is a trivial solution. In particular, we
assume that $\Psi_0(a=0)=0$.

Suppose that $\Psi_0(K,B,c)$ in (\ref{originalsol}) is a tachyon vacuum
solution in the conventional theory. Then,
$\Phi_0(K',B,c)$ in (\ref{solKprime}) is a tachyon vacuum solution in
the theory with $Q'$ for $a>-1/2$, but, in the case of $a=-1/2$,
$\Phi_0(K',c)$ in (\ref{solKprime2}) is a trivial pure gauge solution.

Here, we take $\Psi_a=\Psi_0(a)+\Phi_0$ for $a\ge -1/2$.
We can easily find that $\Psi_a$ is a classical solution in the
conventional theory, namely it satisfies $Q_{\rm B}\Psi_a+\Psi_a^2=0$.
Expanding the string field around $\Psi_a$ in the action, 
we have the kinetic operator
$Q_{\Psi_a}$: $Q_{\Psi_a}A=Q_{\rm B}A+\Psi_aA-(-1)^{|A|}A\Psi_a$ for 
all string field $A$. $Q_{\Psi_a}$ can be written as
\begin{eqnarray}
 Q_{\Psi_a}A=Q'A+\Phi_0 A-(-1)^{|A|}A\Phi_0=Q'_{\Phi_0}A,
\end{eqnarray}
where $Q'$ is the modified BRST operator in (\ref{Qprime}) and
$Q'_{\Phi_0}$ represents the kinetic operator around the solution
$\Phi_0$ in the theory at the identity-based vacuum $\Psi_0(a)$. 
The important point
is that we can construct a homotopy operator for
$Q'_{\Phi_0}(=Q_{\Psi_a})$ for 
$a\ge -1/2$ by use of $K'Bc$ algebra.\footnote{
In the case that $\Psi_0(K,B,c)$ is the Erler-Schnabl 
solution\cite{Erler:2009uj}, e.g.,
we have
$\Phi_0(K',B,c)=\frac{1}{\sqrt{1+K'}}(c+cK'Bc)\frac{1}{\sqrt{1+K'}}$ and
the corresponding homotopy operator is given by a homotopy state:
$\frac{1}{\sqrt{1+K'}}B\frac{1}{\sqrt{1+K'}}$ in the same way as
Ref.~\citen{Inatomi:2012nv}. 
}

Differentiating the equation of motion, $Q_{\rm B}\Psi_a+\Psi_a^2=0$,
with respect to $a$, we find
\begin{eqnarray}
 Q_{\Psi_a}\frac{d}{da}\Psi_a=0.
\end{eqnarray}
Since $Q_{\Psi_a}$ has vanishing cohomology, we have
\begin{eqnarray}
 \frac{d}{da}\Psi_a=Q_{\Psi_a}\Lambda_a,
\label{daPsia}
\end{eqnarray}
for some state $\Lambda_a$.
Integrating (\ref{daPsia}) from $a=0$, we get
\begin{eqnarray}
 \Psi_0(a)+\Phi_0
=\Psi_0(K,B,c)+\int_0^{a}Q_{\Psi_a}\Lambda_a da,
\label{IdequivKBc}
\end{eqnarray}
where we have used the fact that in the case of $a=0$, $\Psi_0(a=0)=0$
and the $K'Bc$ solution is the same as the conventional tachyon vacuum
solution: $\Phi_0(K',B,c)=\Psi_0(K,B,c)$.

From (\ref{IdequivKBc}), we can calculate the gauge invariant overlap
for the identity-based solution:
\begin{eqnarray}
 O_V(\Psi_0(a))=O_V(\Psi_0(K,B,c))-O_V(\Phi_0),
\label{overlapid}
\end{eqnarray}
where we have used the fact that the gauge invariant overlap is 
BRST invariant with respect to $Q_{\Psi_a}$:
$O_V(Q_{\Psi_a}(\cdots))=0$. Noting that the formula (\ref{overlapid})
holds for $a\ge -1/2$, by using the result of the gauge invariant
overlap for $\Phi_0$ in the previous section, the above is evaluated as
\begin{eqnarray}
 O_V(\Psi_0(a))=\left\{
\begin{array}{ll}
 0 & (a>-1/2)\\
&\\
\displaystyle
 \frac{1}{\pi}\left\langle V(i\infty)
	       c(\frac{\pi}{2})\right\rangle_{C_{\pi}} & (a=-1/2) 
\end{array}\right.,
\end{eqnarray}
where we have used the notation in Ref.~\citen{Kishimoto:2013sra}.
Thus, as expected for the identity-based solution, the gauge invariant 
overlap for $a>-1/2$ is equal to that of trivial pure gauge solutions,
and, in the case that  $a=-1/2$, the gauge invariant  overlap agrees
with the result for the tachyon vacuum solution.

As emphasized in Ref.~\citen{Kishimoto:2014qza}, the formula
(\ref{IdequivKBc}) is nothing but a gauge equivalence relation between
$\Psi_0(a)+\Phi_0$ and $\Psi_0(K,B,c)$. In fact, given the
relation (\ref{IdequivKBc}), $\Psi_0(a)+\Phi_0$ can be written as
\begin{eqnarray}
 \Psi_0(a)+\Phi_0=g^{-1}Q_{\rm B}g+g^{-1}\,\Psi_0(K,B,c)\,g,
\end{eqnarray}
where $g$ is given by the path-ordered exponential form,
\begin{eqnarray}
 g={\rm P}\exp\left(\int_0^a \Lambda_a da\right).
\end{eqnarray}
From this gauge equivalence relation, we have
\begin{eqnarray}
 S[\Psi_0(a);Q_{\rm B}]+S[\Phi_0;Q']=S[\Psi_0(K,B,c);Q_{\rm B}].
\end{eqnarray}
From the result for $S[\Phi_0;Q']$ in the previous section and 
for the conventional tachyon vacuum, namely
$S[\Psi_0(K,B,c);Q_{\rm B}]=1/(2\pi^2)$, we finally find that
\begin{eqnarray}
 -S[\Psi_0(a);Q_{\rm B}]=\left\{
\begin{array}{ll}
 0 & (a>-1/2)\\
&\\
\displaystyle
 -\frac{1}{2\pi^2} & (a=-1/2)
\end{array}\right..
\end{eqnarray}
Thus, we have evaluated the vacuum energy density for the identity-based
solution and these results are consistent with our expectation for
the solution.

\section{Concluding remarks
\label{sec:4}
}

We have constructed classical solutions $\Phi_0$
in the theory expanded around the identity-based scalar solution
$\Psi_0$ (\ref{sol}).
We have taken advantage of the $K'Bc$ algebra to calculate observables for
the solution. 
In the case that $\Psi_0$ is trivial pure gauge, the
observables for $\Phi_0$ are equivalent to those for the corresponding
solution $\Psi_0(K,B,c)$ in the original background. 
In the case that $\Psi_0$ is the tachyon vacuum,
they become equal to those for trivial solutions, since the $K'Bc$
algebra is simplified and all the solutions made from $K'$, $B$, 
and $c$ are given as $Q'$-exact states. Finally, we have provided the
gauge equivalence relation between $\Psi_0(a)+\Phi_0$ and
$\Psi_0(K,B,c)$, which is regarded as a new expression for the
identity-based solution.
Thanks to this expression, we have analytically calculated observables
for the identity-based scalar solution whether it corresponds to trivial
pure gauge or tachyon vacuum.

Around the identity-based tachyon vacuum solution, the zeros of
$e^{h(z)}$ on the unit circle play a crucial role in evaluating
observables for 
$\Phi_0$. As seen in Sect.~\ref{sec:2-7}, 
there is no need for these zeros to be at $z=\pm 1$, 
which correspond to open string boundaries.
We note that we can find similar results in the study of homotopy
operators for the BRST operator around the identity-based scalar
solutions,\cite{Inatomi:2011xr,Inatomi:2011an} in which homotopy 
operators exist only if the zeros are on the unit circle.
Here, we should comment on another identity-based solution discussed in
Ref.~\citen{Igarashi:2005wh}, in which $e^{h(z)}$ has higher-order zeros
than the function in this paper. However, similar to the discussion of
homotopy operators in Ref.~\citen{Inatomi:2011xr}, we can obtain the
simplified $K'Bc$ algebra with higher-order zeros and an important point
is the position of the zeros rather than the order.

For the simplest function $h_a(z)$ (\ref{haz}),  
the solution $\Phi_0(K',B,c)$ in the theory around $\Psi_0(a)$ depends
on the parameter $a$.
We find that for $a>-1/2$, $\Phi_0$ can correspond to the tachyon
vacuum but for $a=-1/2$, it becomes a trivial pure gauge
configuration as stated in Sect.~\ref{sec:ObsTV}.
This result is in accordance with the numerical analysis
in Ref.~\citen{Takahashi:2003ppa}, 
where it is observed that in the theory 
around $\Psi_0(a>-\tfrac{1}{2})$,
we can construct a numerical solution whose energy density corresponds
to the negative of the D-brane tension, while the solution continuously
connects to trivially zero as $a$ approaches $-1/2$. We find that the
transition becomes sharp if the truncation level increases.
Accordingly, $\Phi_0$ can be identified as the numerical solution
in~Ref.~\citen{Takahashi:2003ppa} although they belong to different
gauge sectors.
Here, we should note that, for $a=-1/2$, we expect that there exists a
solution whose energy density is the {\it positive} of the D-brane
tension. Such a solution has been constructed  numerically in the Siegel
gauge in Refs.~\citen{Kishimoto:2009nd, Kishimoto:2011zza}.
It should represent the perturbative vacuum where a D-brane
exists.\footnote{N.~Ishibashi pointed out the possibility of
constructing the perturbative vacuum solution
in a private discussion.\cite{Ishibashi}}

\section*{Acknowledgements}
We would like to thank Loriano Bonora and other organizers of the
conference SFT2014 at SISSA, Trieste for the kind hospitality.
We are grateful to Nobuyuki Ishibashi for productive discussions. 
The work of the authors is supported by a JSPS Grant-in-Aid for
Scientific Research (B) (\#24340051). The work of I. K. is supported in
part by a JSPS Grant-in-Aid for Young Scientists (B) (\#25800134).

%

\appendix
\section{Solutions for the differential equation (\ref{deqf})
\label{sec:appendixA}
}

Let us consider solutions of (\ref{deqf}). 
It is sufficient to solve the equation in an annulus including the
unit circle $|z|=1$, because now we look for a regular function $f(z)$
on the circle.

The differential equation (\ref{deqf}) is reducible to the
homogeneous equation,
\begin{eqnarray}
 \frac{dg(z)}{dz}-2i X(z) g(z)=0,~~~~X(z)=-\frac{1}{(1+z^2)e^{h(z)}},
\label{deqg}
\end{eqnarray}
by the variable transformation,
\begin{eqnarray}
 f(z)=i\frac{2g(z)+i}{2g(z)-i}.
\label{fg}
\end{eqnarray}
Solving Eq.~(\ref{deqg}) under the initial condition $f(1)=1$,
the function $f(z)$ is given by
\begin{eqnarray}
 f(z)=-i\,\frac{1+ie^{v(z)} }{1
-ie^{v(z)} },~~~~~v(z)=2i\int_1^z X(z')dz'.
\label{fz1}
\end{eqnarray}

Since the function $X(z)$ has singularity at $z=\pm i$, $v(z)$ becomes
divergent and so the expression (\ref{fz1}) is undefined at the
midpoints.
Here, let us analyze the behavior of $f(z)$ near the midpoints in terms
of series solutions. Suppose that $h(z)$ is holomorphic at $z=i$.
Since $X(z)$ has a single pole at $z=i$, $X(z)$ is expanded into a
Laurent series: 
\begin{eqnarray}
 X(z)=\frac{1}{z-i}\sum_{n=0}^\infty x_n (z-i)^n,
\end{eqnarray}
where the first few coefficients are given by
\begin{eqnarray}
 x_0=\frac{i e^{-h(i)}}{2},~~~
 x_1=-\frac{e^{-h(i)}}{4}(1+2i\partial h(i)),~~~\cdots.
\end{eqnarray}
Using this expansion, we can construct a series
solution for the differential equation (\ref{deqg}):
\begin{eqnarray}
 g(z)=(z-i)^\lambda \sum_{n=0}^\infty A_n (z-i)^n,
\end{eqnarray}
where we find that $A_0\neq 0$, $\lambda=-e^{-h(i)}$, and other $A_n$
are given by a recurrence formula:
\begin{eqnarray}
 A_n=\frac{2i}{n}(x_1 A_{n-1}+x_2A_{n-2}+\cdots+x_n A_0).
\end{eqnarray}
It can be easily seen that this series solution is convergent in a
neighborhood of $z=i$. Since $\lambda=-1$ due to $h(\pm i)=0$ for
the identity-based solution, $f(z)$ is a holomorphic
function near $z=i$:
\begin{eqnarray}
 f(z)=i\frac{2(z-i)g(z)+i(z-i)}{2(z-i)g(z)-i(z-i)}.
\end{eqnarray}
Taking the limit $z\rightarrow i$, we find that $f(i)=i A_0/A_0=i$.
In addition, we have  $f'(i)=-1/A_0\neq 0$.
Similarly, we find that $f(z)$ is holomorphic at $z=-i$, 
and that $f(-i)=-i$ and $f'(-i)\neq 0$.

We have found that the
poles of $X(z)$ at $z=\pm i$ are harmless to solve (\ref{deqf}).
However, if $X(z)$ has poles on the unit circle 
due to zeros of $e^{h(z)}$, it is difficult
to find regular solutions on the unit circle.
Suppose that $X(z)$ is expanded around 
the zero $z_0 (\neq \pm i,|z_0|=1)$ as
\begin{eqnarray}
 X(z)=\frac{x'_0}{z-z_0}+x'_1+\cdots,
\end{eqnarray}
$g(z)$ behaves around $z=z_0$ as
\begin{eqnarray}
 g(z)\sim (z-z_0)^{2ix_0'}\times (\cdots ), 
\end{eqnarray}
where the dots denote a regular function.
Here it is noted that $X(z)$ has poles on the unit circle, but
the residue $x_0'$ is essentially unrestricted
as opposed to the residue $x_0$ at $z=\pm i$.
Therefore, $g(z)$ is not a regular function in general and so
it is impossible to find a regular conformal transformation $f(z)$
if $e^{h(z)}$ has zeros on the unit circle.
Actually, we have seen an example for a singular map in
Sect.~\ref{secexample}.

Now, let us consider (\ref{prop1}) and (\ref{prop2}) for the
solution (\ref{fz1}). 
For $z=e^{i\sigma},~(|z|=1)$, $v(z)$ is written by
\begin{eqnarray}
 v(e^{i\sigma})=\int_0^\sigma \frac{1}{e^{h(e^{i\sigma})}\cos\sigma}
  d\sigma.
\label{vesigma}
\end{eqnarray}
As mentioned in the introduction, the reality condition of $\Psi_0$
implies $(h(z))^*=h(1/z^*)$ and so, for $z=e^{i\sigma}$, $h(z)$ is a
real-valued function. Then, from (\ref{vesigma}), we find that
$v(e^{i\sigma})$ is real-valued. Consequently, from (\ref{fz1}), we find
that $|f(z)|=1$ for $|z|=1$.

For $z=e^{i\sigma}$, we write the phase of $f(z)$ as $\phi(\sigma)$:
\begin{eqnarray}
 \phi(\sigma)=\frac{1}{i}\ln f(e^{i\sigma}).
\end{eqnarray}
Differentiating the phase with respect to $\sigma$, we have
\begin{eqnarray}
 \frac{d\phi(\sigma)}{d\sigma}=\frac{2e^{v(e^{i\sigma})}}{
\{1+e^{2e^{v(e^{i\sigma})}}\}e^{h(e^{i\sigma})}\cos\sigma}.
\end{eqnarray}
Since $v(e^{i\sigma})$ and $h(e^{i\sigma})$ are real, the derivative
is positive for $|\sigma|\leq \pi/2$ and so $\phi(\sigma)$ is a
monotonically increasing function from $-\pi/2$ to $\pi/2$. Hence, we
have proved the properties (\ref{prop1}) and (\ref{prop2}).

Finally, we consider the inversion formula (\ref{prop3}). We note that
the differential equation (\ref{deqf}) has symmetries under the following
transformations:
\begin{eqnarray}
 z &\rightarrow& -\frac{1}{z},\\
 f &\rightarrow& \frac{af+b}{-bf+a},~~~~
(a^2+b^2=1,~~a,b\in{\mathbb C}).
\end{eqnarray}
The first is a ${\mathbb Z}_2$ transformation derived from $h(-1/z)=h(z)$,
which is needed for the
identity-based solution as mentioned in the introduction. The second
transformation forms the group $SO(2,{\mathbb C})$ in which $f=\pm i$ are 
fixed points. Therefore, if a special solution $f(z)$ is known, a
general solution is given by the above $SO(2,{\mathbb C})$ transformation of
$f(z)$. In fact, $SO(2,{\mathbb C})$ has two real parameters and these
correspond to integration constants for the complex first-order
differential equation (\ref{deqf}). Then, since $f(-1/z)$ is also a
solution due to the first symmetry, we find that the relation
\begin{eqnarray}
 f\left(-\frac{1}{z}\right)&=&\frac{af(z)+b}{-bf(z)+a}
\end{eqnarray}
has to hold for some $SO(2,{\mathbb C})$ parameters $a,\,b$.
By performing this transformation twice, we can determine the
parameters as $(a,b)=(1,0)$ or $(0,1)$.
Consequently, since $f(z)$ is holomorphic at $z=i$ and $f'(i)\neq 0$,
$f(z)$ must satisfy the inversion formula (\ref{prop3}).

\end{document}